\documentclass[12pt]{article}
\usepackage{epsfig}

\usepackage{color}
\usepackage[colorlinks,linkcolor=blue,citecolor=red]{hyperref} 

\begin{document}

\title{Two  Component Kaup - Kupershmidt Equation }
\author{Ziemowit Popowicz \\
 University of Wroc\l aw,  Institute of Theoretical Physics\\
 pl.M.Borna 9  50-205 Wroc\l aw Poland\\
 e-mail: ziemek@ift.uni.wroc.pl }
\maketitle

\begin{abstract}
The Kaup - Kupershmidt  equation is generalized to the system of equations in the same manner as  the 
Korteweg - de Vries equation is generalized to the Hirota - Satsuma equation. The Gelfand - Dikii  - Lax 
and Hamiltonian formulation for this generalization is given. 
The same construction is repeated for the 
constrained Kadomtsev - Pietviashvili - Lax operator what leads to   the four component Kaup - Kupershmidt 
equation. 
The modified version of the two component 
Kaup - Kupershmidt equation is presented and analysed.
\end{abstract}

\section*{Introduction.} 
Large classes of nonlinear partial differential equations are integrable by the inverse spectral transform method 
and its modifications  \cite{1,2}.
It is well known that most of the integrable partial differential equations ,
\begin{equation}
u_t=F(t,x,u,u_x,u_{xx},...)
\end{equation}
admit so called  Lax representation 
\begin{equation}
\frac{\partial L}{\partial t} = \Big [A,L\Big ],
\end{equation}
and hence the inverse scattering method is applicable.

We shall consider the case where the Lax operator is a differential operator 
\begin{equation}
L=\partial^m+u_{m-2}\partial^{m-2} + ... + u_0,
\end{equation}
where $u_i, i=0,1, ... m-2$ are functions of $x,t$.
Then the equation (2) gives us the Gelfand - Dikii system  where 
 $ A= L^{n/m} $  is a pseudodifferential series of the form $ L^{n/m} = \sum_{-\infty}^{i=n}v_i\partial^i$
and $ L^{n/m}_{\geq 0}  = \sum_{i=0}^{n}v_i\partial^i$. 

Quite different systems of equations could be obtained considering the Kadomtsev - Pietviashvili hierarchy within
Sato's approach \cite{3,4}. In this case the Lax operator is spanned by infinitely many fields 
\begin{equation}
L_{KP}=\partial + u_1\partial^{-1}+u_2\partial^{-2} + ....
\end{equation}
with the following Lax pair representation
\begin{equation}
\frac{\partial L}{\partial t} = \Big [(L^N)_{\geq 0},L\Big ],
\end{equation}

Both these hierarchies describe  a large classes of nonlinear partial differential  equations. 
In order to find some interesting equations, in these hierarchies,  sometime we need to   
to apply the  reduction procedure in which some functions are described in terms of  other functions 
used in the Lax operator. 
We have no the unique prescription how to carry out  such procedure at the moment. 
Kupershmidt \cite{5} has noticed, that certain invariance of the partial differential nonlinear equations, 
can be extracted from the Lax operator. This observation allowed him to put some constraints on the 
functions appearing 
in the Lax operator. This procedure, is called now the Kupershmidt reduction \cite{1}.

In this paper we would like to consider some specific reduction of the Gelfand - Dikii Lax operator 
in which Lax operator can be factorized as the product of two Lax operators. This idea follows from the 
observation that the product of two Lax operators \cite{6} of the  Korteweg - de Vries equations 
\begin{equation}
L=(\partial^2 + u)(\partial^2 + v)
\end{equation}
creates the whole hierarchy of equations with  the following 
Lax pair representation 
\begin{equation}
\frac{\partial L}{\partial t_n} = 8\Big [(L^{(2n+1)/4})_{\geq 0},L\Big ],
\end{equation}
where $n=0,1,2,...$  and factor 8 was chosen in such a way that to normalize the higher term in 
the equation.
For $n=1$ we have   Hirota - Satsuma equation \cite{7}
\begin{eqnarray}
\frac{\partial u}{\partial t_1} &=&\Big (-u_{xxx}+3v_{xxx}-6u_xu+6vu_x+12v_xu\Big ), \\ \nonumber
\frac{\partial v}{\partial t_1} &=&\Big (-v_{xxx}+3u_{xxx}-6v_xv+6v_xu+12vu_x\Big ) ,
\end{eqnarray}
while for $n=2$ 
\begin{eqnarray}
\frac{\partial u}{\partial t_2} &=&\Big (-3u_{xxxxx}-15u_{xxx}u-15u_{xx}u_x-15u_xu^2+5v_{xxxxx} + \\ \nonumber
&& \quad  25v_{xxx}u + 5v_{xxx}v+25v_{xx}u_x+15v_{xx}v_x + 15v_xu_{xx}+ \\ \nonumber 
&& \quad  20v_xu^2 + 20v_xvu +5v^2u_x+5vu_{xxx} + 30vu_xu \quad \Big )/4, \\ \nonumber
\frac{\partial v}{\partial t_2} &=&\Big (5u_{xxxxx} + 5u_{xxx}u +15u_{xx}u_x - 3v_{xxxxx} + 5v_{xxx}u - \\ \nonumber 
&& \quad  15v_{xxx}v +15v_{xx}u_x -15v_{xx}v_x +25v_xu_{xx} +5v_xu^2 -\\ \nonumber
&& \quad 15v_xv^2 +30v_xvu + 20v^2u_x + 25vu_{xxx} + 20vu_xu \quad \Big )/4 ,
\end{eqnarray}
 
Let us notice that both these equations  could be rewritten in the hamiltonian form as 
\begin{equation}
\left(
\begin{array}{c}
u \\ v
\end{array}
\right)_{t_{n}} = J
\left (
\begin{array}{c} \frac{\delta H_n}{\delta u} \\ \frac{\delta H_n}{\delta v} \end{array} \right )=
\left(
\begin{array}{cc}
-\frac{1}{2} \partial^3  -2u\partial -u_x & 0  \\  0 & -\frac{1}{2} \partial^3 - 2v\partial - v_x \end{array}\right) 
\left (
\begin{array}{c} \frac{\delta H_n}{\delta u} \\ \frac{\delta H_n}{\delta v} \end{array} \right ), 
\end{equation}
where $n=1,2$ and 
\begin{eqnarray}
H_1 &=& \int dx \ Res( L^{3/2}) =\int dx \ (u^2+v^2-6uv) , \\ \nonumber 
H_2 &=& \int dx \ Res(L^{5/2}) = \int dx \ ((3u_{xx}+10v_{xx})u - u^3 - 3v_{xx}v - v^3 + 5vu(v+u) )
\end{eqnarray}
and $Res$ denotes the coefficient standing in the $\partial^{-1}$ term.

Recently it was  showed in \cite{8} that the similar construction could be carried out for the Harry Dym equation
which leads  to the  system of interacting equations.  
However the Lax operator for the Harry Dym equation does not belong to the Gelfand - Dikii system.

Both these equations could be considered either as the extensions of known equations or as the 
reduction of the Lax pair representations.
Indeed  Lax operator (6)  could be considered as the admissible reduction of the fourth-order
 Gelfand - Dikii - Lax operator 
\begin{equation}
L=\partial^4+f_2\partial^2 + f_1\partial + f_0,
\end{equation}
where 
\begin{equation}
\quad f_2 = u+v, \quad f_1=2v_x, \quad f_0=v_{xx}+vu.
\end{equation}

We  would like now  to repeat the similar construction for the Boussinesq type  Lax operators. 
We choose third order Lax operator of the form
\begin{equation}
L=\partial^3 + u\partial + \lambda u_x
\end{equation}
where at the moment $\lambda $ is a free parameter.

This Lax operator generate the whole hierarchy of equations and the first nontrivial equation  starts from 
the fifth flow 
\begin{equation}
\frac{\partial L}{\partial t_5} = 9 \Big [(L^{(5/3})_{\geq 0},L\Big ],
\end{equation}
of the form 
\begin{equation}
u_t=\Big ( -u_{4x} -5u_{xx}u + 15\lambda(\lambda -1) u_x^2 - \frac{5}{3}u^3 \Big )_x
\end{equation}
only when $\lambda = \frac{1}{2}, 1 , 0$. 
Notice that the factor 9 was  chosen in such a way that  to normalize the higher terms in the equation. 

For $\lambda =\frac{1}{2}$ we have Kaup - Kupershmidt hierarchy \cite{9,10} while for  $\lambda = 1$ or $\lambda = 0$ 
we obtain  Sawada - Kotera  hierarchy \cite{11}.  
Both these equations are hamiltonian equations where  
\begin{equation}
u_t=\big (c\partial^3 + \frac{1}{15}(\partial u + u\partial )\Big )\frac{\delta H}{\delta u}
\end{equation}
where 
\begin{equation}
H_1=\int dx \quad \Big ( 3(3\lambda^2 - 3\lambda +1)u_x^2 - 5u^3 \Big ) 
\end{equation}
and $c=\frac{2}{15}$ for $\lambda = \frac{1}{2}$ or $c=\frac{1}{15}$ for $\lambda = 1$ 
or $\lambda = 0$

Now we are prepare to consider new Lax operator as the product of two different 
Lax operators of the  Boussinesq type 
\begin{equation}
L:=(\partial^3 + v\partial + \lambda v_x)(\partial^3+(u-v)\partial + \lambda (u_x - v_x))
\end{equation}
The consistent hierarchy could be obtained only for $\lambda = \frac{1}{2}$ and first two 
nontrivial flows are
\begin{equation}
\frac{\partial L}{\partial t_n} = 9 \Big [(L^{(n/6})_{\geq 0},L\Big ],
\end{equation}
give us 
\begin{eqnarray}
v_{t_3} &=& \frac{9}{2}\Big (u_{xxx} -2v_{xxx} + \frac{1}{2} v_xu -3v_xv + vu_x\Big ) \\ \nonumber
u_{t_3} &=& \frac{9}{2}\Big ( -\frac{3}{4} u^2 -3v^2 + 3uv \Big )_x
\end{eqnarray}

\begin{eqnarray}
v_{t_5} &=& \Big ( -5u_{xxxx} + 9v_{xxxx}  - \frac{5}{2}u_{xx}u - \frac{5}{2} u_x^2 + 15v_{xx}v   
 +\frac{15}{4}v_x^2 + \\ \nonumber 
&& \quad \frac{5}{2}v^3 - \frac{5}{2}vu_{xx} - \frac{5}{8}vu^2 \Big )_x - 
 \frac{5}{2}vu_{xxx} -\frac{5}{4}vu_xu \\ \nonumber
u_{t_5} &=& \Big ( - u_{xxxx}  + 5u_{xx}u + \frac{35}{24}u^3 -15v_{xx}u +30v_{xx}v +\frac{15}{2}v_x^2 - \\ \nonumber 
&&  \quad \frac{15}{2}v_vu_x +\frac{15}{2} u^2v - 15vu_x{xx} -\frac{15}{2}vu^2 \Big )_x
\end{eqnarray}
The last system of the equations is our two component generalized Kaup - Kupershmidt equation. This system
cannot be reduced  to the system of equations (9) by the linear transformation.

Interestingly this two - component generalization have been considered first time in [4] where authors 
investigated  the so called constrained Kadomtsev - Pietviashvilli (KP) hirerarchy.
The constrained KP hierarchy is obtained from the usual KP hierarchy as
\begin{equation}
L^{N}_{KP}=(L^{N}_{KP})_{\geq 0} + \Psi\partial^{-1}\Phi
\end{equation}
with $L_{KP}$ defined by (4).
Then the equations (21,22) could be obtained  choosing 
\begin{equation}
L_{KP}=\partial^3 +\frac{1}{2} u\partial + \frac{1}{4}u_x + \frac{1}{16}(2v-u)\partial^{-1}(2v-u)
\end{equation}

In contrast to the usual Kaup - Kupershmidt hierarchy, which starts from the fifth flow, our hierarchy 
begin from the third flow. 
Notice that our Lax operator as well as the equations  allows the reduction to the standard 
Kaup - Kupershmidt  Lax operator or equations when $u=2v$. 

Both these systems are hamiltonians where 
\begin{equation}
\left(
\begin{array}{c}
u \\ v
\end{array}
\right)_{t_{n}} =  J \ \frac{\delta H_n}{\delta v} =
\frac{1}{216} \left(
\begin{array}{cc}
4\partial^3 +\partial u + u \partial  & 2\partial^3 + \partial v + v \partial   \\  
2\partial^3  + \partial v + v\partial  & 2\partial^3 + \partial v + v \partial \end{array}\right) 
\left (
\begin{array}{c} \frac{\delta H_n}{\delta u} \\ \frac{\delta H_n}{\delta v} \end{array} \right ) 
\end{equation}
and
\begin{eqnarray}
H_3 &=& \int dx   Res(L^{3/6}) = 54 \int dx  \Big ( 4uv - 4v^2 - u^2 \Big ) \\ \nonumber 
H_5 &=& \int dx Res(L^{5/6}) = \int dx  \Big ( 7u^3 + 24u_{xx}u - (108v_{xx}- 36vu)(v - u)  \Big )
\end{eqnarray}
By straightforward calculations it is easy to show that the hamiltonian operator $J$  satisfy  the Jacobi 
identity.

Let us now consider the following Miura transformation 
\begin{equation}
u = a_x \quad , \quad v = b_x - \frac{1}{4}b^2.
\end{equation}
where $a,b$ are functions of $x$ and $t$.
It is easy to show that this transforms the systems of equations 
\begin{eqnarray}
a_{t_{3}} &=& \frac{1}{16} \Big ( -12a_x^2 - 48b_x^2 + 48b_xa_x + 24b_xb^2-3b^4 -12b^2a_x \Big ) \\ \nonumber
b_{t_{3}} &=& \frac{1}{4}  \Big ( 4a_{xx} - 8b_{xx} + b^3 + 2ba_x \Big )_x
\end{eqnarray}
\begin{eqnarray}
a_{t_{5}} &=& \frac{1}{96} \Big ( -96a_{xxxxx} +   480a_{xxx}a_x + 140a_x^3 + 1440b_{xxx}(2b_x-a_x) + \\ \nonumber
&& 720b_{xx}(-a_{xx} - 3b_xb + \frac{1}{2}b^3 ) + 45b^4a_x + 360b^2a_{xxx} +180b^2a_x^2 \\ \nonumber
&& 360b_x(-4b_x^2+4b_xa_x + \frac{3}{2}b_xb^2-4a_{xxx} -b^2a_x +ba_{xx}) \Big ) \\ \nonumber
b_{t_{5}} &=& \frac{1}{32}  \Big (-160a_{xxxx} - 80a_{xx}a_x  + 288b_{xxxx} -240b_{xx}b_x -120b_{xx}b^2 \\ \nonumber
&& -12b_x^2b +3b^5 -80ba_{xxx} -20ba_x^2  \Big )_x
\end{eqnarray}
to the systems (21) or (22) respectively.  

Notice that the  equations (28) describe system of two interacting fields of the  modified Korteweg - de Vries 
type.  This system of equations  does not belong to the class of the interacting fields considered by 
Foursov \cite{12}.
Foursov has classified all integrable systems of  two interacting modified KdV - type equations which 
could  be reduced to the symmetrical form 
\begin{equation}
u_t = F[u,v] \quad , \quad v_t =F[v,u] ,
\end{equation}   
where $F[u,v]=F[u,u_x,u_{xx}, .. v,v_x,v_{xx} ...]$ denotes differential polynomial function of two variables.
However our  system of equations (24) cannot be reduced to the symmetrical form by the linear transformation . 

Interestingly the system (28) collapses when $u=2v$. Indeed the condition $u=2v$ is equivalent with the 
assumption that 
\begin{equation}
a_x = 2b_x - \frac{1}{2}b^2
\end{equation}
and therefore we have  $a_{t_3}=0$. The system of equation (29) reduces when $u=2v$  to the modified version 
of the Kaup - Kupershmidt equation
\begin{equation}
b_t=\frac{1}{16} \Big ( -16b_{xxxx} - 40b_{xx}b_x +20 b_{xx}b^2+20b_x^2b-b^5 \Big )_x
\end{equation}
Our equations (28) and (29) are hamiltonians equations where 
\begin{equation}
\left(
\begin{array}{c}
a \\ b
\end{array}
\right)_{t_{n}} = {\cal D}
\left (
\begin{array}{c} \frac{\delta H_n}{\delta a} \\ \frac{\delta H_n}{\delta b} \end{array} \right )=
\frac{1}{2} \left(
\begin{array}{cc}
-4\partial  - (\partial^{-1}a_x - a_x\partial^{-1})  & 
-2\partial - \partial^{-1}b_x + b 
  \\  -2\partial -  b - b_x\partial^{-1}
& - 2\partial  \end{array}\right) 
\left (
\begin{array}{c} \frac{\delta H_n}{\delta a} \\ \frac{\delta H_n}{\delta b} \end{array} \right ), 
\end{equation}
where $n=3,5$ and 
\begin{eqnarray}
H_3 &=& \int dx \quad \Big ( \frac{1}{2}a_{xx}a - 2b_{xx}a + 2b_{xx}b +b_xba-\frac{1}{8}b^4 \Big ) \\ \nonumber
H_5 &=& \int dx \quad \Big ( 24a_{xxxx}a + 14a_{xx}a_xa - 108b_{xxxx}(a -b) +54b_{xxx}b_xb +\\ \nonumber 
&& \quad b_{xx}(234b_xa-36a_xa -108b_xb-18b^2a) +b_x^2(27b^2-36ba) + \\ \nonumber 
&& \quad  b_x(bb^3a -36a_{xx}a + 18ba_aa) + 9b^2a_{xx}a \Big ) 
\end{eqnarray}
It is easy to check that the operator $\cal D$ is the Hamiltonian operator. Indeed it is enough to 
notice that under the Miura transformation (27) this operator transforms to the $\hat J = \cal F \cal D  \cal F^{\star}$ 
where $\cal F$ is the Freche derivative of Miura transformation  and $\star$ denotes the 
hermitian conjugation.
\begin{equation} 
\hat J=\left( 
\begin{array}{cc}
\partial  &  0 \\ 0
& - \partial - \frac{1}{2} b  \end{array}\right) 
\end{equation}

Let us apply finally  the factorization procedure directly to constrained Kadomtsev - Petviashvili - Lax 
operator.  We consider therefore two different Lax operators 
\begin{eqnarray}
L_{1} &=& \partial^3 + v\partial + \frac{1}{2}v_x + h \partial^{-1} h, \\ \nonumber 
L_{2} &=& \partial^3 + (u-v)\partial + \frac{1}{2}(u_x - v_x) + g \partial^{-1} g, 
\end{eqnarray}
and construct new Lax operator as
\begin{equation} 
L=L_1L_2. 
\end{equation}
This Lax operator generate the integrable hierarchy of four interacting fields. The first 
nontrivial equations are 

\begin{eqnarray}
u_{t_3} &=& \frac{9}{2}\Big (6g^2+6h^2 -\frac{3}{2}u^2+6vu -6v^2 \Big )_x \\ \nonumber 
v_{t_3} &=& \frac{9}{2}(12hh_x+2u_{xxx}-4v_{xxx}  +v_xu-6v_xv+2vu_x) \\ \nonumber 
g_{t_3} &=& \frac{9}{2} (2g_{xxx}-u_xg+ug_x+3v_xg ) \\ \nonumber 
h_{t_3} &=& \frac{9}{2}(2h_{xxx}+2u_xh+uh_x-3v_xh ),
\end{eqnarray}

\begin{eqnarray}
u_{t_5} &=& \Big ( 60g_{xx}g + 15g_x^2 + 60h_{xx}h  +15h_{x}^2  -u_{xxxx} +5u_{xx}u  +\frac{35}{24}u^3 \\ \nonumber 
&& -\frac{15}{2}ug^2 +\frac{75}{2}uh^2-15v_{xx}u +30v_{xx}v+\frac{15}{2}v_{x}^2 -\frac{15}{2}v_{x}u_{x} 
 +\frac{15}{2}v^2u \\ \nonumber 
 && +45vg^2 -45vh^2  -15vu_{xx}-\frac{15}{2}vu^2 \Big)_{x} , \\ \nonumber 
v_{t_5} &=& 30g_{xxx}g+90g_{xx}g_x+30h_{xxx}h-5u_{xxxxx}  -\frac{5}{2}u_{xxx}u-\frac{15}{2}u_{xx}u_{x} 
+30u_{x}h^2+ \\ \nonumber 
&& 45uh_{x}h  +9v_{xxxxx}+15v_{xxx}v+\frac{45}{2}v_{xx}v_{x}+\frac{15}{2}v_{x}g^2  -\frac{75}{2}v_{x}h^2-
\frac{5}{2}v_{x}u_{xx}  \\ \nonumber 
&& -\frac{5}{8}v_{x}u^2+\frac{15}{2}60v_{x}v^2  +
30vg_xg-60vh_{x}h 
 \quad  -5vu_{xxx}-\frac{5}{2}vu_{x}u \\ \nonumber 
g_{t_5} &=& 9g_{xxxxx}-\frac{15}{2}g_xg^2+30h_{x}hg+\frac{15}{2}h^2g_x +\frac{5}{2}u_{xxx}g-
\frac{5}{2}u_{xx}g_x +\frac{5}{4}u_{x}ug -\frac{5}{8}u^2g_x \\ \nonumber 
&&  +\frac{15}{2}ug_{xxx}+\frac{45}{2}v_{xx}g_x+\frac{45}{2}v_{x}g_{xx} 
  +\frac{15}{4}v_{x}ug-\frac{15}{2}v_{x}vg-\frac{15}{2}v^2g_x+\frac{15}{2}vug_x \\ \nonumber 
h_{t_5} &=& 9h_{xxxxx}+\frac{15}{2}h_{x}g^2-\frac{15}{2}h_{x}h^2+30hg_xg  +\frac{5}{2}u_{xxx}h+20u_{xx}h_{x} \\ \nonumber 
&& +\frac{45}{2}u_{x}h_{xx}  -\frac{5}{2}u_{x}uh-\frac{5}{8}u^2h_{x}+\frac{15}{2}uh_{xxx}-
\frac{45}{2}v_{xx}h_{x} \\ \nonumber 
&& -\frac{45}{2}v_{x}h_{xx}+\frac{15}{4}v_{x}uh-\frac{15}{2}v_{x}vh-\frac{15}{2}v^2h_{x} +
\frac{15}{2}vu_{x}h+\frac{15}{2}vuh_{x}
\end{eqnarray}
The last system of equations could be considered as the four - component generalized 
Kaup - Kupershmidt equation. This equation reduces to the two - component 
Kaup - Kupershmidt equation when $g=h=0$.


\begin{thebibliography}{99}

\bibitem{1} M. B\l aszak {\it Multi - Hamiltonian Theory of Dynamical Systems} Springer-Verlag 1998.

\bibitem{2} I. M. Gelfand, L. A. Dikii Funct.Anal.Appl. {\bf 10} (1976) 250.

\bibitem{3} M. Sato Y. Sato {\it Lecture Notes in Num. Appl. Anal} {\bf 5} 259 (1982)

\bibitem{4} W. Oevel and W. Strampp Comm.Math.Phys. {\bf 157} 51 (1993).

\bibitem{5} B. Kupershmidt Commu.Math.Phys. {\bf 99} (1988) 51.

\bibitem{6} R. Dodd and A. Fordy  Phys.Lett. {\bf 89A} (1982) 168.

\bibitem{7} R. Hirota and J. Satsuma Phys.Lett. {\bf 85A} (1981) 407.

\bibitem{8} Z. Popowicz Phys.Lett. {\bf A 317} (2003) 260.

\bibitem{9} A. P. Fordy and J. Gibbons J.Math.Phys. {\bf 21} (1980) 2508.

\bibitem{10} D. J.Kaup Stud.Appl. Math. {\bf 62} (1980) 189.

\bibitem{11} K. Sawada and T. Kotera Prog. Theor. Phys. {\bf 51} (1974) 1255.

\bibitem{12} M. Foursov J. Math. Phys. {\bf 41} (200) 6173.

\end{thebibliography}
\end{document}